\documentclass[letter,traditabstract]{aa} 
\usepackage{graphicx}
\usepackage{natbib}
\usepackage{amssymb}
\usepackage{amsmath}
\usepackage{url}
\usepackage[section]{placeins}
\usepackage{multirow}
\usepackage{txfonts}
\newcommand{\change}[1]{#1}          

\begin{document}
   \title{First hyperfine resolved far-infrared OH spectrum from a star-forming region\thanks{\textit{Herschel} is an ESA space observatory with science instruments provided by European-led Principal Investigator consortia and with important participation from NASA.}}

\author{S.~F.~Wampfler \inst{\ref{inst1}} 
\and S.~Bruderer \inst{\ref{inst2}}
\and L.~E.~Kristensen \inst{\ref{inst3}}
\and L.~Chavarr{\'{\i}}a\inst{\ref{inst4},\ref{inst5}}
\and E.~A.~Bergin \inst{\ref{inst6}}
\and A.~O.~Benz  \inst{\ref{inst1}} 
\and E.~F.~van~Dishoeck \inst{\ref{inst2},\ref{inst3}}
\and G.~J.~Herczeg \inst{\ref{inst2}}
\and F.~F.~S.~van~der~Tak \inst{\ref{inst7},\ref{inst8}}
\and J.~R.~Goicoechea \inst{\ref{inst9}}
\and S.~D.~Doty \inst{\ref{inst10}}
\and F.~Herpin \inst{\ref{inst4},\ref{inst5}}
}

\institute{
Institute for Astronomy, ETH Zurich, 8093 Zurich, Switzerland\label{inst1}
\and
Max Planck Institut f\"{u}r Extraterrestrische Physik, Giessenbachstrasse 1, 85748 Garching, Germany\label{inst2}
\and
Leiden Observatory, Leiden University, PO Box 9513, 2300 RA Leiden, The Netherlands\label{inst3}
\and
Universit\'{e} de Bordeaux, Observatoire Aquitain des Sciences de l'Univers, F-33271 Floirac Cedex, France \label{inst4}
\and
CNRS, UMR 5804, Laboratoire d'Astrophysique de Bordeaux, 2 rue de l'Observatoire, BP 89, F-33271 Floirac Cedex, France\label{inst5}
\and
Department of Astronomy, The University of Michigan, 500 Church Street, Ann Arbor, MI 48109-1042, USA\label{inst6}
\and
SRON Netherlands Institute for Space Research, PO Box 800, 9700 AV, Groningen, The Netherlands\label{inst7}
\and
Kapteyn Astronomical Institute, University of Groningen, PO Box 800, 9700 AV, Groningen, The Netherlands\label{inst8}
\and
Departamento de Astrof\'{\i}sica, Centro de Astrobiolog\'{\i}a (CSIC-INTA), 28850, Madrid, Spain\label{inst9}
\and
Department of Physics and Astronomy, Denison University, Granville, OH, 43023, USA\label{inst10}
}


\date{accepted May 25, 2011} \titlerunning{Velocity resolved OH observation of W3~IRS~5}


\def\placefigureHIFISpectrum{
\begin{figure}
 \centering
 \resizebox{0.95\hsize}{!}{\includegraphics[angle=270]{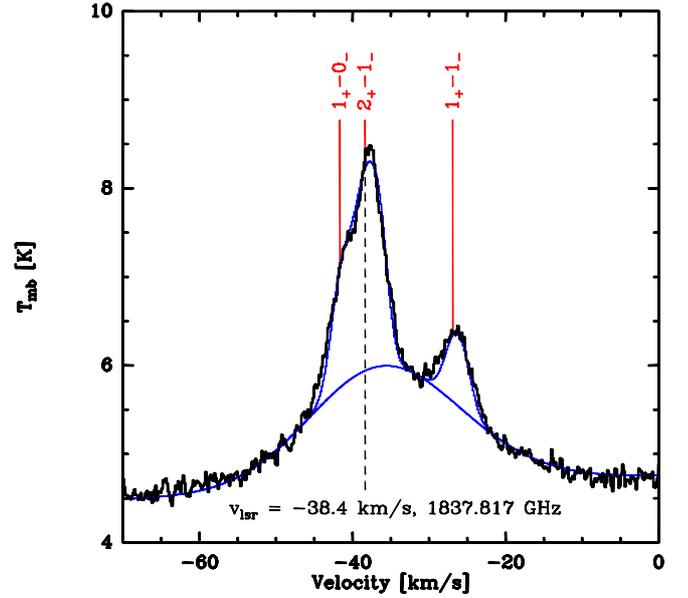}}
 \caption{HIFI spectrum of the OH triplet at 1837.8~GHz with half the dual-sideband continuum. The expected positions of the lines and the source velocity are labeled. The blue lines show the best fit from the slab models and the outflow component separately. }
 \label{fig:hifi_spectrum}
\end{figure}
}

\def\placefigureEnvelopeModels{
\begin{figure}
 \centering
 \resizebox{0.95\hsize}{!}{\includegraphics[angle=270]{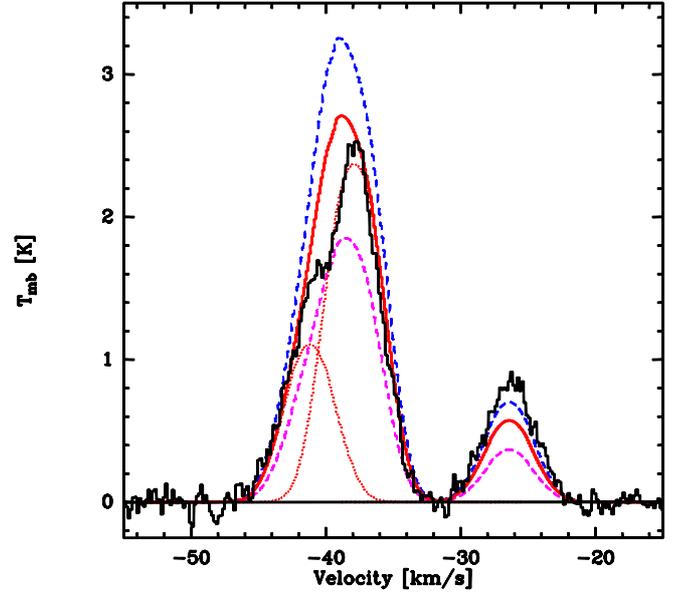}}
 \caption{Resulting HIFI spectrum after subtraction of the best fit outflow component. Overplotted are the spherical envelope models for constant OH abundances of $1 \times 10^{-8}$ (blue dashed), $8 \times 10^{-9}$ (red solid with individual components red dotted) and $5 \times 10^{-9}$ (pink dashed). Hyperfine components were simply added (see text).}
 \label{fig:envelope_models}
\end{figure}
}

\def\placefigureLineWidthComparison{
\begin{figure}[h]
 \centering
 \resizebox{0.95\hsize}{!}{\includegraphics[angle=270]{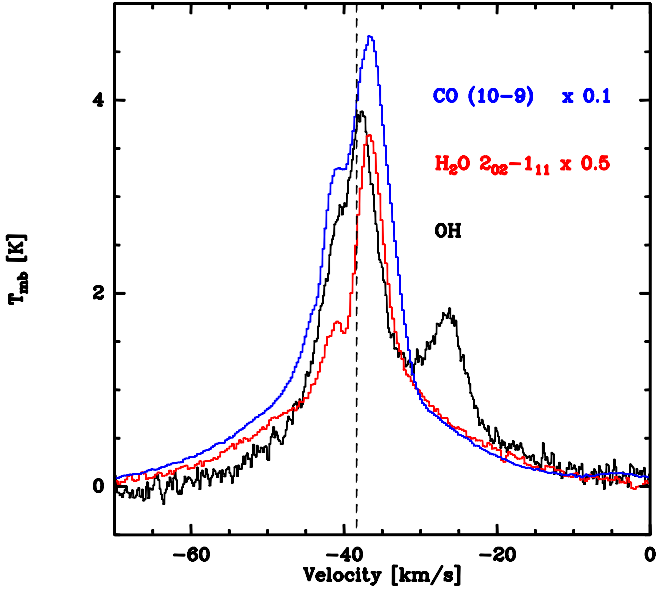}}
 \caption{Comparison of the broad component of OH (black), H$_2$O \mbox{$2_{0 2}- 1_{1 1}$ (red)}, and CO(10-9) (blue) observed with HIFI. The black dashed line indicates the source velocity ($\varv_{\mathrm{lsr}} = -38.4~\mathrm{km}~\mathrm{s}^{-1}$).}
 \label{fig:line_widths}
\end{figure}
}

\def\placefigureChiSquare{
\begin{figure}[htb]
 \centering
 \resizebox{0.8\hsize}{!}{\includegraphics[angle=0]{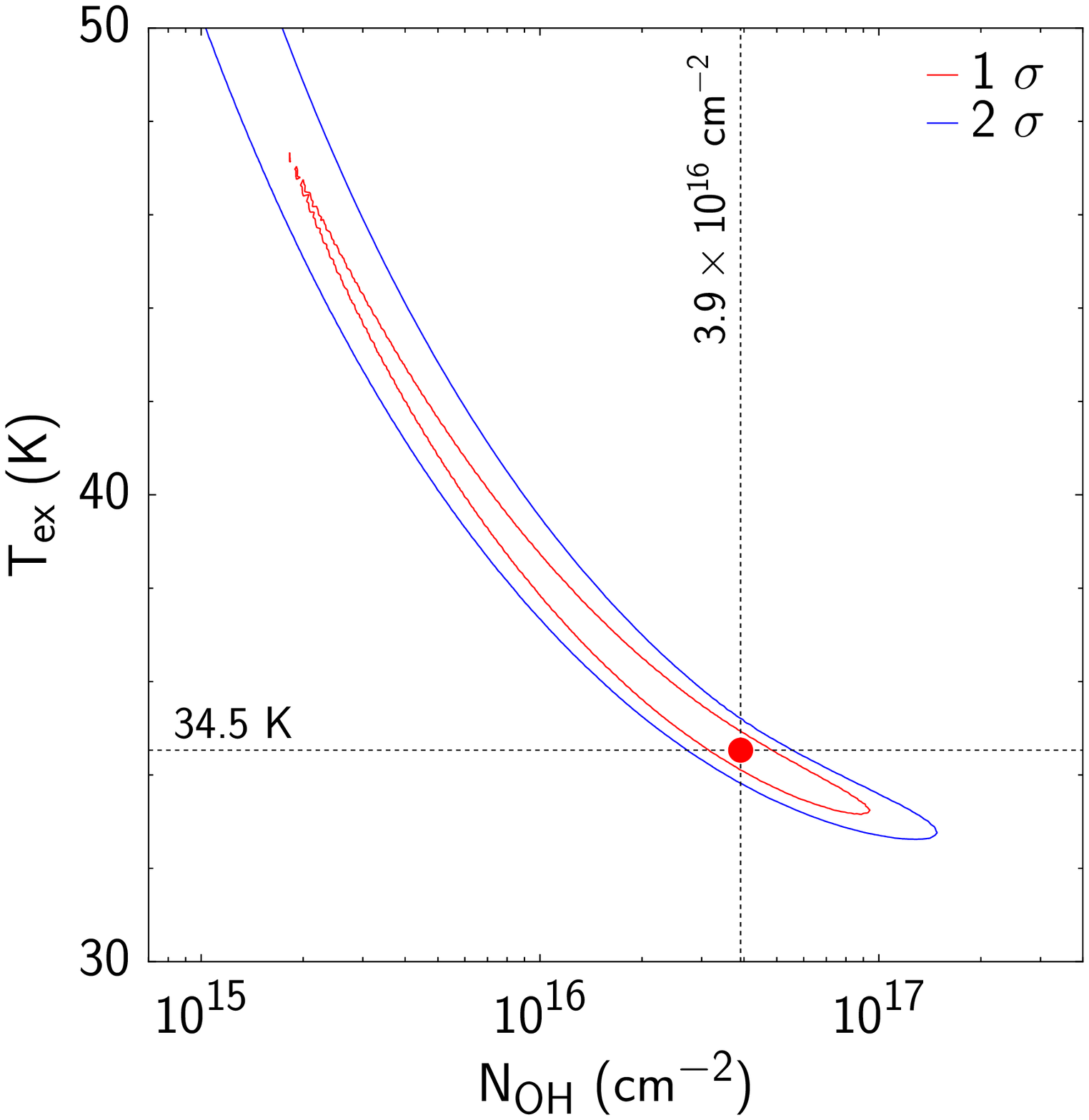}}
 \caption{$1\sigma$ and $2\sigma$ contours for the fit of the narrow (envelope) component with slab models. The best fit is indicated by the red dot.}
 \label{fig:chi2env}
\end{figure}
}

\def\placetableMoldata{
\begin{table}
\caption{Molecular data of the observed OH transitions between the ${}^2\Pi_{1/2}$ 3/2 and 1/2 excited states.}
\begin{center}
\begin{tabular}{l l l l l l}
\hline 
\hline
Transition        & Frequency & $A_{\mathrm{ul}}$ & $g_u$ & $g_l$ & Shift \\
F, P              & [GHz]     & [s$^{-1}$]        &       &       & [km~s$^{-1}$] \\ 
\hline
$1+ \rightarrow 1-$ & 1837.7466 & 2.1(-2)         & 3     & 3     & \phantom{-}11.5\\
$2+ \rightarrow 1-$ & 1837.8168 & 6.4(-2)         & 5     & 3     & \phantom{-0}0.0\\
$1+ \rightarrow 0-$ & 1837.8370 & 4.3(-2)         & 3     & 1     & \phantom{0}-3.3\\
 \hline
\end{tabular}
\end{center}
\label{tab:moldata}
\begin{flushleft}
\footnotesize{$A(B) \equiv A \times 10^B$. The last column denotes the velocity shift relative to the component with the largest Einstein A coefficient.}
\end{flushleft}
\end{table}
}

\def\placetableFit{\begin{table}[tb]
\caption{\change{Beam-averaged} column density $N_{\rm OH}$, excitation temperature $T_{\rm ex}$, line width $\Delta \varv$, and line position $\varv_{\rm lsr}$ of the best fit slab model. 
}
\centering
\begin{tabular}{llllll}
\hline
\hline
   & $N_{\rm OH}$            & $T_{\rm ex}$    & $\Delta \varv$ & $\varv_{\rm lsr}$ & fit range \\
   & [cm$^{-2}$]             & [K]             & [km/s]         & [km/s]            & [km/s]\\
\hline
O  &           $\ge$7.2(13)  & 270.0$^a$       &           21.9 & -35.7             & [-60,-47.5], [-22,-10] \\
E  & \phantom{$\ge$}3.9(16)  & \phantom{0}34.5 & \phantom{0}3.3 & -37.7             & [-43,-35], [-30,-23] \\
\hline
\end{tabular}
\begin{flushleft} 
\footnotesize{${}^{a}$ fixed (see text). $A(B) \equiv A \times 10^B$. O = outflow, E = envelope.}
\end{flushleft}
\label{tab:fit}
\end{table}
}

\def\placetableIntIntensity{
\begin{table}[htb]
\caption{Gaussian fit results to the OH line components using the velocity range [-100,30] and first order baselines. \change{The $\varv-\varv_{\rm lsr}$ velocity scale is given relative to the laboratory frequency of the \mbox{$2+ \rightarrow 1-$} component \mbox{($\varv_\mathrm{lsr} = -38.4~\mathrm{km}~\mathrm{s}^{-1} = 1837.817~\mathrm{GHz}$)}.}}
\begin{center}
\begin{tabular}{l l l l l}
\hline 
\hline
Component                    & $\int{T_{\mathrm{mb}} d\varv}$ & $T_{\mathrm{peak}}$ & $\Delta \varv$       & $\varv-\varv_{\rm lsr}$      \\
                             & [K~km~s$^{-1}$]                & [K]                 & [km~s$^{-1}$]        & [km~s$^{-1}$] \\ 
\hline
Envelope $1+ \rightarrow 1-$ & \phantom{0}$3.9\pm 0.4$        & 0.8                 & \phantom{0}4.5$^{a}$ & \phantom{-}11.9 $\pm$ 0.2\\
Envelope $2+ \rightarrow 1-$ &           $11.7\pm 0.6$        & 2.4                 & \phantom{0}4.5$^{a}$ & \phantom{-1}0.8 $\pm$ 0.1\\
Envelope $1+ \rightarrow 0-$ & \phantom{0}$5.7\pm 0.5$        & 1.2                 & \phantom{0}4.5$^{a}$ & \phantom{1}-3.3 $\pm$ 0.2\\
Outflow (blended)            &           $36.7\pm 1.3$        & 0.8                 & $27.4 \pm 1.0$       & \phantom{-1}4.1 $\pm$ 0.4\\
\hline
\end{tabular}
\end{center}
\label{tab:int_intensity}
\begin{flushleft}
\footnotesize{${}^{a}$ Parameter fixed. The given errors are only fitting errors and do not include the calibration uncertainty.}
\end{flushleft}
\end{table}
}

 
\abstract
   {OH is an important molecule in the H$_2$O chemistry and the cooling budget of star-forming regions. The \change{goal of the} \textit{Herschel} key program `Water in Star-forming regions with \textit{Herschel}' (WISH) \change{is to study} H$_2$O and related \change{species} during protostellar evolution. Our \change{aim in this letter} is to assess the origin of the OH emission from star-forming regions and constrain the properties of the emitting gas. High-resolution observations of the OH ${}^2\Pi_{1/2}~J = 3/2-1/2 $ triplet at $1837.8~\mathrm{GHz}$ ($163.1~\mu$m) towards the high-mass star-forming region W3~IRS~5 with the Heterodyne Instrument for the Far-Infrared (HIFI) on \textit{Herschel} \change{reveal} the first hyperfine velocity-resolved OH far-infrared spectrum of a star-forming region. The line profile of the OH emission \change{shows} two components: a narrow component (FWHM $\approx 4-5~\mathrm{km}~\mathrm{s}^{-1}$) with \change{partially} resolved hyperfine structure resides on top of a broad (FWHM $\approx 30~\mathrm{km}~\mathrm{s}^{-1}$) component. The narrow emission \change{agrees well with results from radiative transfer calculations of a spherical envelope model for W3~IRS~5 with} a constant OH abundance of $x_{\mathrm{OH}} \approx 8 \times 10^{-9}$. Comparison with H$_2$O yields OH/H$_2$O abundance ratios of around $10^{-3}$ for $T \gtrsim 100~\mathrm{K}$ and around unity for $T \lesssim 100~\mathrm{K}$, consistent with the current picture of the dense cloud chemistry with freeze-out and photodesorption. The broad component is attributed to outflow emission. An abundance ratio of OH/H$_2$O $\gtrsim 0.028$ in the outflow is derived from comparison with results of water line modeling. This ratio can be explained by a fast J-type shock or a slower UV-irradiated C-type shock.}


\maketitle

\section{Introduction} 
Newly-formed stars inject large amounts of energy into the ambient interstellar material through shocks and radiation, heating the surrounding gas and dust. In these warm regions, H$_2$O becomes one of the most abundant gas-phase molecules \change{because} water ice evaporates from the grain mantles and gas\change{-}phase formation routes become available. The hydroxyl radical (OH) is closely linked to both the H$_2$O formation and destruction through \mbox{OH + H$_2$ $\Leftrightarrow$ H$_2$O + H} as well as a byproduct of the H$_2$O photodissociation process in the presence of UV photons. A significant fraction of the gas cooling occurs through line emission in the far-infrared, including lines of $[$\ion{O}{i}$]$, $[$\ion{C}{ii}$]$, CO, H$_2$O, and OH \citep[][]{Giannini01,vanKempen10}. 

Observations of the OH far-infrared transitions first became possible with the Kuiper Airborne Observatory \citep[e.g.][and references therein]{Storey81,Betz89,Melnick90}. OH was frequently detected in star-forming regions with ISO \citep[e.g.][]{Giannini01,Goicoechea06} and with PACS \citep[][]{Wampfler10} on board the \textit{Herschel} Space Observatory \citep{Pilbratt10}. \change{Previously, OH was mainly observed through maser lines at cm wavelengths \citep[][for models see e.g. \citealt{Cesaroni91}]{Gaume87}.} Observations carried out with ISO and PACS lack the spectral resolution needed to distinguish between OH emission from the quiescent envelope and the outflow, since no information on the line shape is obtained. Furthermore, the hyperfine structure of the OH lines remaines unresolved. The Heterodyne Instrument for the Far-Infrared \citep[HIFI, ][]{DeGraauw10} on \textit{Herschel} offers an adequate spectral resolution to resolve the 1837.8~GHz triplet of OH. Thus, optical depths and line widths to constrain the temperature and density of the emitting gas can be derived. \change{A blended OH triplet was previously detected with HIFI from Orion KL \citep{Crockett10}.}

This letter presents the first hyperfine velocity resolved observations of OH obtained using HIFI from the \change{well-studied} high-mass star-forming region W3~IRS~5 located at a distance of 2~kpc \citep{Hachisuka06} with a total luminosity of $\sim 10^5~\mathrm{L}_{\sun}$ \citep[e.g. ][]{Helmich97,Boonman03,vanderTak05,Chavarria10,Benz10}. 

\section{Observations and data reduction}

The OH triplet at 1837.747, 1837.817 and 1837.837~GHz from W3~IRS~5 (\mbox{$\alpha_{2000}=02^{\mathrm{h}} 25^{\mathrm{m}} 40\fs60$}, \mbox{$\delta_{2000}=+62\degr 05\arcmin 51\farcs0$}) was observed with HIFI on \textit{Herschel}. The observations were carried out as part of the `Water in Star-Forming Regions with Herschel' (WISH) key program \citep{vanDishoeck11}. W3~IRS~5 was observed on July 29$^{\mathrm{th}}$ 2010 (obsid 1342201666) in dual beam switch mode with an on-source integration time of 17~min. The beam size (HPBW) at 1837~GHz is about $12\arcsec$. The average system temperature was 1245~K. The wide band spectrometer (WBS) offers a nominal resolution of 1.1~MHz, corresponding to a velocity resolution of about $0.18~\mathrm{km}~\mathrm{s}^{-1}$ at 1837~GHz. \change{The OH triplet at 1834.7~GHz was not observed.}

HIFI data were reduced using the Herschel interactive processing environment \citep[HIPE v4.0.0, ][]{Ott10} and the GILDAS-CLASS\footnote{http://www.iram.fr/IRAMFR/GILDAS} software. The H and V polarizations were combined, reaching $T_{\mathrm{rms}} \approx 0.1~\mathrm{K}$ . We \change{subtracted} a first order polynomial from the spectra and calibrated to $T_{\mathrm{mb}}$ scale using a main beam efficiency of 0.70. Molecular data (Table \ref{tab:moldata}) are taken from the Leiden atomic and molecular database LAMDA\footnote{http://www.strw.leidenuniv.nl/$\sim$moldata/} \citep{Schoier05} with data from the JPL catalog\footnote{http://spec.jpl.nasa.gov} \citep{Pickett98} and collisional rate coefficients by \citet{Offer94}.

The calibration uncertainty of band 7b is currently estimated to \change{$\lesssim 32\%$} and is mainly caused by the unknown side band gain ratio. The uncertainty reduces to about 10\% for a ratio close to unity. The errors on the integrated intensities are only fitting errors and do not include the calibration uncertainty. The velocity calibration \change {is uncertain by} $\sim 0.2~\mathrm{km}~\mathrm{s}^{-1}$ \change{and} could \change{partially} explain the \change{$\sim 0.5~\mathrm{km}~\mathrm{s}^{-1}$} offsets of the line peaks from the source velocity of $\varv_{\mathrm{lsr}} = -38.4~\mathrm{km}~\mathrm{s}^{-1}$ \change{\citep{vanderTak00b}}.

\section{Results}

HIFI clearly detects the OH triplet at around 1837.8~GHz ($163.1~\mu$m) towards W3~IRS~5 (Fig. \ref{fig:hifi_spectrum}). For the first time, the hyperfine components of the line triplet are spectrally resolved.

Two components dominate the line shape: the line profile shows narrow components residing on top of a broad emission feature, similar to that found by \citet{Chavarria10} for H$_2$O lines in W3~IRS~5 and \citet{Kristensen10} in low-mass YSOs. The narrow components with full widths at half maximum \change{(FWHM)} of \mbox{$\sim 4-5~\mathrm{km}~\mathrm{s}^{-1}$} are centered close to the positions expected from the hyperfine pattern. The lines are split by $20.2~\mathrm{MHz}$ ($3.3~\mathrm{km}~\mathrm{s}^{-1}$) towards the blue and $70.2~\mathrm{MHz}$ ($11.5~\mathrm{km}~\mathrm{s}^{-1}$) towards the red relative to the middle line. Therefore, the \mbox{($F=2+ \rightarrow1-$;} Tab. \ref{tab:moldata}) and ($F=1+ \rightarrow0-$) transitions are blended, while the ($F=1+ \rightarrow1-$) transition is resolved.
The underlying broad component with a \change{FWHM} of $\sim 30~\mathrm{km}~\mathrm{s}^{-1}$ consists of the three blended hyperfine components, each with $\Delta {\rm \varv} \sim 20$ km s$^{-1}$. The hyperfine pattern is unresolved in the broad component.

\change{Gaussian fits to the narrow components are presented in the online appendix (Sec. \ref{sec:gaussfit}). The derived hyperfine intensity ratios (1:$3\pm0.5$:$1.5\pm0.3$) deviate from the prediction of LTE in the optically thin limit ($\sim A_{\mathrm{ul}} g_\mathrm{u}$, 1:5:2).} 
Pumping by far-infrared dust continuum radiation, line overlap, and optical depth may \change{all} contribute to the observed hyperfine anomaly. Modeling results (Sec. \ref{sec:slab_models}) indicate that at least the $2+ \rightarrow 1-$ line is optically thick. Based on comparison with spherical envelope models (Sec. \ref{sec:envelope_models}), we attribute the narrow components to OH emission from the hot parts of the envelope.

The broad component in OH is most likely associated with outflow emission. Its integrated intensity is around $37~\mathrm{K}~\mathrm{km}~\mathrm{s}^{-1}$ (Sec. \ref{sec:gaussfit}). A similar component can also be identified in the line profiles of other species like CO and H$_2$O (Sec. \ref{sec:linewidth}) from recent Herschel results and ground-based observations \citep[e.g. ][]{Hasegawa94,Boonman03,Chavarria10}. 

\placetableMoldata

\placefigureHIFISpectrum

\subsection{Slab Models} \label{sec:slab_models}

As a first modeling step to derive OH column densities, the slab method outlined in \citet[][appendix B]{Bruderer10} is used. This method calculates the molecular spectrum from two slabs -- one representing the envelope and one the outflow -- in front of a continuum source, determined from the observed continuum flux. The slabs are assumed to cover the entire continuum source. The free parameters describing each slab are the OH column density $N_{\mathrm{OH}}$, the excitation temperature $T_{\mathrm{ex}}$ \change{(assumed to be equal for all hyperfine transitions)}, the line width $\Delta {\mathrm{\varv}}$, and the position of the line center ${\mathrm{\varv}}_{\mathrm{lsr}}$. Overlap between different hyperfine components of OH is taken into account. To constrain the free parameters ($N_{\rm OH}$, $T_{\rm ex}$, $\Delta {\rm \varv}$ and ${\rm \varv}_{\rm lsr}$), the $\chi^2$ between observation and model is minimized using an uncertainty of $T_{\mathrm{rms}} \sim 0.1~\mathrm{K}$. 

The resolved hyperfine triplet structure of the narrow (envelope) component allows us to constrain the column density and the excitation temperature of the emitting gas simultaneously. Results of the best fitting slab models for the narrow component with a reduced $\chi^2 = 1.2$ are presented in Table \ref{tab:fit}. A peak line opacity of $\tau=1.8$ is reached ($2+ \rightarrow 1-$ transition) and the $1+ \rightarrow 0-$ and  $1+ \rightarrow 1-$ peaks have optical depths of $\tau = 0.8$ and $\tau = 0.4$, respectively. 
Figure \ref{fig:chi2env} of the online appendix shows the $1\sigma$ and $2\sigma$ contours for the column density and excitation temperature. The OH column densities within the $1\sigma$ interval range from $N_{\rm OH} = 2 \times 10^{15}-1 \times 10^{17}$ cm$^{-2}$, depending on $T_{\mathrm{ex}}$, and thus vary by almost two orders of magnitude. 

A broad (outflow) component can be clearly identified in the spectrum, but its hyperfine components are not resolved. Thus, it is not possible to determine an excitation temperature of the broad emission from the spectrum. Assuming $T_{\rm ex} = 100$ K, a column density of $N_{\rm OH} = 2.0 \times 10^{14}$ cm$^{-2}$ is obtained by fitting the line wings only. A lower limit on the OH column density of $N_{\rm OH} \geq 7.2 \times 10^{13}$ cm$^{-2}$ is derived by assuming that the excitation temperature is equal to the upper level energy (270 K) of the transition. Both fits yield a line width of $\Delta \varv = 21.9$ km s$^{-1}$ and a line position $\varv_{\rm lsr}=-35.7~\mathrm{km}~ \mathrm{s}^{-1}$. The line optical depth is \mbox{$\tau < 0.01$} for these excitation temperatures. An optical depth \change{of} \mbox{$\tau = 1$} is reached for $T_{\rm ex} < 34$ K and no good fit can be obtained for $T_{\rm ex} \lesssim 32$ K. In the following, $T_{\rm ex} = 270~\mathrm{K}$ is assumed.

\placetableFit

To convert the column density into an OH abundance in the outflow, the ${}^{12}$CO(3-2) observation of \citet{Hasegawa94} \change{was used} to derive $N_{\mathrm{H}_2} = 1.3 \times 10^{21}~\mathrm{cm}^{-2}$ with RADEX \citep{vanderTak07}, assuming a CO/H$_2$ abundance ratio of $10^{-4}$, a density of $10^5~\mathrm{cm}^{-3}$, and a temperature of $60~\mathrm{K}$, as derived in their paper. This H$_2$ column density converts the lower limit of $N_{\rm OH} \geq 7.2 \times 10^{13}$ cm$^{-2}$ into a lower limit on the OH abundance of $x_{\mathrm{OH}} = N_{\mathrm{OH}}/N_{\mathrm{H}_2} \geq 5.5 \times 10^{-8}$.

\subsection{Spherical model of the envelope} \label{sec:envelope_models}
In addition to the slab modeling, the narrow emission component is compared to the results from full radiative transfer models using the `RATRAN' code \citep{Hogerheijde00}. The physical structure of W3~IRS~5 is taken from \citet{vanderTak00b} with a power-law density profile ranging from $n =  10^{5}-10^{8}~\mathrm{cm}^{-3}$ and temperatures $50 \lesssim T \lesssim 950~\mathrm{K}$ within the \textit{Herschel} beam at $1837~\mathrm{GHz}$. Dust and gas temperatures are assumed to be equal. \change{F}ar-infrared excitation by the dust continuum for $T_{\mathrm{dust}}$ is \change{included}. We also adopt their distance (2.2~kpc) for consistency. Dust opacities based on grains with thin ice mantles are assumed \citep[][ table 1, column 5]{Ossenkopf94}. The molecular abundance is assumed to be constant throughout the envelope, because this is the simplest structure yielding a good fit to the line profiles. Figure \ref{fig:envelope_models} shows the comparison between model and data, where the best fitting slab model for the broad component (cf. Sec. \ref{sec:slab_models}) is subtracted from the observation, leaving only the narrow component for a simpler comparison.

\placefigureEnvelopeModels

Models assuming an OH abundance of $x_{\mathrm{OH}} = (0.5-1) \times 10^{-8}$ \change{agree} reasonably \change{well} with the observed narrow components. RATRAN does not treat line overlap and overlap effects can therefore not be treated accurately for optically thick lines. Thus, we have simply added the intensities of the components as would be appropriate in the optically thin limit. 

The spherical envelope models are also consistent with the $79, 84$ and $119~\mu\mathrm{m}$ lines being in absorption, as observed in the unpublished spectral scan obtained with \change{PACS} on \textit{Herschel}. \change{The $79$ and $119~\mu\mathrm{m}$ transitions are connected to the ground ${}^2\Pi_{3/2}$ level and therefore prone to absorption.} A quantitative analysis is not possible because the central spatial pixel of the detector is saturated. 

\section{Discussion}

Insight into the water chemistry can be gained by comparing OH and H$_2$O abundances, because these species are linked through the OH + H$_2$ $\Leftrightarrow$ H$_2$O + H reactions. \citet{Chavarria10} identify broad and narrow components in H$_2$O line profiles from W3~IRS~5 similar to OH. OH/H$_2$O abundance ratios can be derived separately for gas of the envelope and the outflow.

From the spherical non-LTE envelope models, which do not consider overlap of hyperfine components, we find an OH abundance of $x_{\mathrm{OH}} \approx 8 \times 10^{-9}$ for W3~IRS~5. New calculations of the H$_2$O abundance $x_{\mathrm{H}_2\mathrm{O}}$ in the envelope, based on the work by \citet{Chavarria10} but \change{recalculated} using the same physical structure as for OH, yield $x_{\mathrm{H}_2\mathrm{O}} = 10^{-5}$ for $T \geq 100~\mathrm{K}$ and $x_{\mathrm{H}_2\mathrm{O}} = 10^{-8}$ for $T < 100~\mathrm{K}$. This allows us to estimate the OH/H$_2$O abundance ratio\change{s}: H$_2$O is about three orders of magnitude more abundant in the inner envelope ($T \geq 100~\mathrm{K}$), while the OH/H$_2$O ratio is around unity in the outer part ($T < 100~\mathrm{K}$). The same conclusion is reached in the best fit H$_2$O abundance structure of \citet{Boonman03} to ISO and SWAS data, which was derived with the same physical model as adopted here.

The inferred OH/H$_2$O ratios \change{in} the envelope are consistent with the current picture of the water chemistry in dense clouds 
\citep[for detailed discussion of processes see][]{Hollenbach09,vanDishoeck11}. Different paths can increase the gas-phase H$_2$O abundance in the inner part of protostellar envelopes. At $T \ge 100~\mathrm{K}$, H$_2$O starts to evaporate from the ice mantles of dust grains. When temperatures of $ T \gtrsim 230~\mathrm{K}$ are reached, which are typical for the innermost parts of high-mass protostellar envelopes, H$_2$O can be rapidly formed in the gas phase through \mbox{OH + H$_2$ $\Rightarrow$ H$_2$O + H}. In this case, OH is only a transient step of the H$_2$O formation process and thus OH/H$_2$O $\ll 1$ in the absence of UV photons. In the outer envelope, low-temperature gas-phase chemistry provides $x_{\mathrm{OH}} \approx 10^{-8}$ \citep[][Fig.~11]{Doty02}. At the outer edge, H$_2$O and OH can be released into the gas phase by photodesorption from grain mantles. \citet{Oberg09} and \citet{Andersson08} find from laboratory data and theoretical calculations that roughly equal amounts of OH and H$_2$O are released at low temperatures. Depending on \change{the} optical depth of the lines, this outer layer may dominate the emission.

The slab model method outlined in Sec. \ref{sec:slab_models} is also used to estimate the H$_2$O abundance in the outflow component. Availability of \change{HIFI} para-H$_2$O $2_{0,2}-1_{1,1}$ \citep[from ][]{Chavarria10} and unpublished $2_{1,1}-2_{0,2}$ (Chavarria et al. in prep.) and $3_{3,1}-4_{0,4}$ data allow us to constrain $N_{\mathrm{p-H}_2\mathrm{O}}$ and the para-H$_2$O excitation temperature in the outflow simultaneously. We find $N_{\mathrm{p-H}_2\mathrm{O}} = 6.4 \times 10^{14}$ cm$^{-2}$ and thus $N_{\mathrm{H}_2\mathrm{O}} = 2.6 \times 10^{15}$ cm$^{-2}$ for an assumed ortho- to para-H$_2$O ratio of 3:1. The water abundance is calculated to be $2.1 \times 10^{-6}$ with $N_{\mathrm{H}_2} = 1.25 \times 10^{21}~\mathrm{cm}^{-2}$. Under the assumption that the OH and H$_2$O emission arise from the same gas, we can calculate a lower limit of 0.028 on the OH/H$_2$O abundance ratio in the outflow \change{of W3~IRS~5}. This limit is consistent with the emission originating in either a fast ($\varv > 60~\mathrm{km}~\mathrm{s}^{-1}$), dissociative J-type shock \citep{Neufeld89} or a slower UV-irradiated C-type shock \citep[$\varv \sim$20--30 km s$^{-1}$,][]{Wardle99}. Standard C-type shocks underproduce the observed column density ratio by one order of magnitude or more (Kristensen et al. in prep.). It is not possible to distinguish between the two types of shocks with the current data and availability of model results.

\section{Conclusions}
The OH lines at $1837.8~\mathrm{GHz}$ from W3~IRS~5 detected with HIFI consist of two components: a broad component, where the triplet components are blended because of the large individual line widths, and a narrow component with velocity resolved hyperfine structure on top. These results indicate that OH emission observed at a lower spectral resolution, e.g. with PACS, can be a blend of envelope and outflow contributions. \change{Low-mass sources reach lower temperatures than the high-mass counterparts at the projected distance of the Herschel beam. From RATRAN models of low-mass sources we thus find a lower excitation, resulting in a weaker envelope contribution to the OH $1837~\mathrm{GHz}$ lines than from W3~IRS~5, consistent with the results for HH 46 in \citet{Wampfler10}.}
Because the broad component also appears in the line profiles of other species from W3~IRS~5, in particular those of CO and H$_2$O, the outflow is the most likely origin of the emission. Comparison with H$_2$O yields a lower limit on the OH/H$_2$O abundance ratio of 0.028, consistent with an origin from a fast J-type shock or a slower UV-irradiated C-type shock. 
The narrow emission \change{agrees well} with spherical envelope models with \change{a constant} OH abundance \change{of} $x_{\mathrm{OH}} \approx 8 \times 10^{-9}$. Comparison with new H$_2$O results based on \citet{Chavarria10} give OH/H$_2$O $\approx 1$ for $T < 100~\mathrm{K}$ and OH/H$_2$O $\approx 10^{-3}$ at $T \gtrsim 100~\mathrm{K}$, \change{consistent} with the current picture of the dense cloud chemistry with freeze-out.

\bibliographystyle{aa}
\bibliography{mybib}

\begin{acknowledgements}
The authors are grateful to an U.A.~Y{\i}ld{\i}z for maintaining the WISH database and to C.~Dedes, T.~Giannini, G.~Melnick, and R.~Visser for useful discussions. \change{We thank the anonymous referee for constructive comments.}
The work on star formation at ETH Zurich is partially funded by the Swiss National Science Foundation (grant nr. 200020-113556). This program is made possible thanks to the Swiss HIFI guaranteed time program.
HIFI has been designed and built by a consortium of institutes and university departments from across Europe, Canada and the United States under the leadership of SRON Netherlands Institute for Space
Research, Groningen, The Netherlands and with major contributions from Germany, France and the US.
Consortium members are: Canada: CSA, U.Waterloo; France: CESR, LAB, LERMA, IRAM; Germany:
KOSMA, MPIfR, MPS; Ireland, NUI Maynooth; Italy: ASI, IFSI-INAF, Osservatorio Astrofisico di Arcetri-
INAF; Netherlands: SRON, TUD; Poland: CAMK, CBK; Spain: Observatorio Astronómico Nacional (IGN),
Centro de Astrobiología (CSIC-INTA). Sweden: Chalmers University of Technology - MC2, RSS \& GARD;
Onsala Space Observatory; Swedish National Space Board, Stockholm University - Stockholm Observatory;
Switzerland: ETH Zurich, FHNW; USA: Caltech, JPL, NHSC.
\end{acknowledgements}

\Online

\begin{appendix}
\section{Gaussian fitting} \label{sec:gaussfit}
\placetableIntIntensity

\section{Line width comparison} \label{sec:linewidth}
The comparison of OH and H$_2$O column densities in the outflow is based on the assumption that the emission arises from the same gas. Figure \ref{fig:line_widths} illustrated the similar widths of the broad components of OH ($\Delta \varv \approx 27~\mathrm{km}~\mathrm{s}^{-1}$), CO ($\Delta \varv \approx 29~\mathrm{km}~\mathrm{s}^{-1}$), and H$_2$O ($\Delta \varv \approx 26~\mathrm{km}~\mathrm{s}^{-1}$). The broad component of OH is a blend of three hyperfine components with individual widths of $\Delta \varv \approx 22~\mathrm{km}~\mathrm{s}^{-1}$, derived from the best fit slab model (Sec. \ref{sec:slab_models}). 
\placefigureLineWidthComparison

\section{Slab modeling details}
\change{The modeling of the OH line spectrum (cf. Sec. \ref{sec:slab_models}) is carried out with the slab model code presented in appendix B of \citet{Bruderer10}. Both OH line components (narrow/envelope and broad/outflow) are represented by a slab in front of a continuum source. The continuum temperature $T_\mathrm{cont}$ is derived from the observed single-sideband continuum value and the source is assumed to be fully covered by both slabs. No geometry is included, except that the outflow slab is placed in front of the envelope slab. Each slab has four free parameters per line: the OH column density $N_{\mathrm{OH}}$, the excitation temperature $T_{\mathrm{ex}}$, the line width $\Delta \varv$, and the position (in velocity space) of the line center $\varv_{\mathrm{lsr}}$, but the excitation temperature is assumed to be the same for all hyperfine transitions. The normalized level populations of the upper ($x_\mathrm{u}$) and lower level ($x_\mathrm{l}$) are therefore determined by the Boltzmann distribution at $T_{\mathrm{ex}}$,
\begin{equation}
\frac{x_\mathrm{u}}{x_\mathrm{l}} =  \frac{g_\mathrm{u}}{g_\mathrm{l}} \exp \left( - \frac{h \nu_0}{k_\mathrm{B}T_{\mathrm{ex}}} \right)
\end{equation}
with the statistical weights $g_\mathrm{u}$ and $g_\mathrm{l}$ of the upper and lower level, respectively, Boltzmann's constant $k_\mathrm{B}$, Planck's constant $h$, and the frequency $\nu_0$ of the transition. The radiation temperature $T\left( \nu \right)$ is derived from the solution of the radiative transfer equation, using the Rayleigh-Jeans approximation, as
\begin{eqnarray}
T \left( \nu \right) &=& T_\mathrm{cont} \left( \nu \right) \exp \left(-\tau^\mathrm{env}\right) \exp \left( -\tau^\mathrm{ofl} \right) \nonumber \\
& & + \frac{c^2}{2 \nu_0 k_\mathrm{B}} B_{\nu_0} \left( T^\mathrm{env}_\mathrm{ex} \right) \left[1 - \exp \left(-\tau^\mathrm{env}  \right) \right] \nonumber \\
& & + \frac{c^2}{2 \nu_0 k_\mathrm{B}} B_{\nu_0} \left( T^\mathrm{ofl}_\mathrm{ex} \right) \exp \left( -\tau^\mathrm{env} \right)\left[1 - \exp \left(- \tau^\mathrm{ofl}  \right) \right]
\end{eqnarray}
with $c$ being the speed of light, $B_{\nu_0}$ the Planck function, $\tau^\mathrm{env}$ the optical depth of the envelope layer, and  $\tau^\mathrm{ofl}$ the optical depth of the outflow layer. The line optical depth of every slab is the sum of the contributions of all $M$ hyperfine components and can be calculated from 
\begin{equation}
\tau \left( \nu \right) = N_\mathrm{OH} \sum_{j = 1}^{M} \frac{c^2}{8 \pi \left( \nu^j\right)^2} A^j_\mathrm{ul} \left( x^{j}_\mathrm{l} \frac{g_\mathrm{u}}{g_\mathrm{l}} -x^{j}_\mathrm{u} \right) \varphi^{j}\left( \nu \right)
\end{equation}
where $A_\mathrm{ul}$ is the Einstein coefficient and $\varphi\left( \nu \right)$ the normalized line profile function of the transition, assumed to be a Gaussian of width $\Delta \varv$ centered at $\varv_{\mathrm{lsr}}$. This approach takes line overlap into account.
}
\placefigureChiSquare
\end{appendix}

\end{document}